\documentclass[manuscript]{aastex}

\slugcomment{To appear in The Astronomical Journal}

\shorttitle{Radio HH Object in W75N}
\shortauthors{}

\begin{document}


\title{A bright radio HH object with large proper motions in the massive  
star-forming region W75N}


\author{Carlos Carrasco-Gonz\'alez\altaffilmark{1,2}, Luis F. Rodr\'\i guez\altaffilmark{2},
Jos\'e M. Torrelles\altaffilmark{3}, Guillem~Anglada\altaffilmark{1}, and Omaira 
Gonz\'alez-Mart\'\i n\altaffilmark{4}}





\altaffiltext{1}{Instituto de Astrof\'\i sica de Andaluc\'\i a (CSIC), 
Camino Bajo de Hu\'etor 50, E-18008 Granada, Spain}
\altaffiltext{2}{Centro de Radioastronom\'\i a y Astrof\'\i sica (UNAM), Apartado 
Postal 3-72 (Xangari), 58089 Morelia, Michoac\'an, M\'exico}
\altaffiltext{3}{Instituto de Ciencias del Espacio (CSIC/IEEC)-UB, Universitat de
Barcelona, Mart\'{\i} i Franqu\`{e}s 1, E-08028 Barcelona, Spain}
\altaffiltext{4}{Physics Department, University of Crete, P. O. Box 2208, 71003 Heraklion, 
Crete, Greece}


\begin{abstract}
We analyze radio continuum and line observations from the archives of the Very Large Array,
as well as X-ray observations from the \emph{Chandra} archive of the region of
massive star formation W75N.
Five radio continuum sources are detected: VLA 1, VLA 2, VLA 3, Bc, 
and VLA 4. VLA 3 appears to be a radio 
jet; we detect J=1-0, v=0 SiO emission towards it, probably tracing
the inner parts of a
molecular outflow.
The radio continuum source Bc, previously believed
to be tracing an independent star, is found to exhibit 
important changes in total flux density, morphology, and position.
These results
suggest that source Bc is actually a radio Herbig-Haro object, one of the
brightest known, powered by the VLA~3 jet source. 
VLA 4 is a new radio continuum component, located
a few arcsec to the south of
the group of previously known radio sources.
Strong and broad (1,1) and (2,2) ammonia emission
is detected from the region containing the radio sources
VLA~1, VLA~2, and VLA~3. 
Finally, the 2-10 keV emission seen in the \emph{Chandra}/ACIS image shows
two regions that could be the termination shocks
of the outflows from the multiple sources observed in W75N.
 
\end{abstract}





\section{Introduction}

The compact, thermal (i. e. free-free) radio sources found at centimeter wavelengths in 
regions of star formation can have different natures. Some of them are
ultracompact or hypercompact HII regions, photoionized by an embedded 
hot luminous star.
Other sources are clumps of gas or even circumstellar disks
that are being externally ionized by a nearby star, such as the
Orion proplyds (e. g. O'Dell \& Wong 1996; Zapata et al. 2004)
and the bright-rimmed clouds (e.g. Carrasco-Gonzalez 
et al. 2006).
An additional class of sources is formed by the
thermal jets, collimated outflows that emanate from young stars and 
whose ionization is most probably maintained by the interaction of the
moving gas with the surrounding medium (Eisloffel et al. 2000). Finally, there are also
some examples of ``radio Herbig-Haro'' objects, obscured knots of gas that are
being collisionally ionized by the shock produced by a collimated outflow
(e. g. Curiel et al. 1993). 
There are also compact non-thermal sources, of which the
most common are the young low-mass
stars with active magnetospheres that produce gyrosynchrotron emission
(e.g. Feigelson \& Montmerle 1999). There are
also emission regions where fast shocks may produce synchrotron emission
(e.g. Henriksen et al. 1991; Garay et al. 1996).

To advance in the understanding of the nature of these compact radio sources
it is necessary to have good quality data that allows the observer
to establish the angular size, the morphology, the spectral index,
the time variability, the polarization,
and the presence of proper motions in them. 
Only a handful of the known star-forming regions have been studied carefully
enough to clearly establish the nature of its compact radio sources. 

The massive star-forming region W75N
is part of the Cygnus X complex of dense molecular clouds. 
Its distance is estimated to be 2 kpc (Fischer et al. 1985).
Haschick et al. (1981) detected at 6 cm a
source that was interpreted as an ultracompact HII (UC HII) region named W75N(B), 
that later was resolved into three small 
diameter radio continuum sources: Ba, 
Bb and Bc in the high angular resolution ($0\rlap.{''}5$)
observations of Hunter et al. (1994).
Two of these sources, Ba and Bb, were also detected at 1.3 cm 
(with a $0\rlap.{''}1$ 
resolution) by Torrelles et al. (1997) (who 
named them as VLA 1 and VLA 3) along with another fainter and more compact source,
located in between VLA 1 and VLA 3, and named as VLA 2. 
These compact sources were believed to be UC HII regions
in an early phase, based on the presence of H$_2$O 
and/or OH masers (Baart et al. 1986, Hunter et al. 1994, Torrelles et al. 1997) in 
its close vicinity. 

At radio wavelengths, VLA 1 has a structure elongated at a PA of $\sim 43^\circ$
(Torrelles et al. 1997),
approximately in the direction of the observed high-velocity three pc-scale 
bipolar outflow (Hunter et al. 1994; 
Shepherd et al. 2003).
However, Shepherd (2001) had pointed that VLA 1
does not appear to
contribute significantly to the energetics of the large-scale outflow.
VLA 2 shows a symmetric roundish shape. A detailed 
Very Long Baseline Interferometry (VLBI)
study of the water maser cluster morphology and kinematics in the vicinity of these 
two radio sources by Torrelles et al. (2003) showed remarkably different outflow 
ejection geometries among them (VLA 1 has a collimated, jetlike outflow,
while VLA 2 has a shell outflow expanding in multiple directions). This 
is a surprising result given that both objects 
are separated by only 1400 AU and share the same environment. VLA 3 was
found to have deconvolved dimensions of $0\rlap.{''}09 \times 0\rlap.{''}04$
$(PA = 152^\circ)$ at 1.3 cm by Torrelles et al. (1997).
This source is also detected at mid-infrared
wavelengths, with a bolometric luminosity of ~ 750~L$_{\odot}$ (Persi,
Tapia, \& Smith 2006). The physical parameters 
of the ionized gas in VLA 1, VLA 2, VLA 3, and Bc, as well as the more evolved W75N(A) 
are all consistent with each being powered by a ZAMS star of spectral type between 
B0 and B2 (Shepherd et al. 2004). The present evidence is consistent with
an independent exciting source of this spectral type for each radio continuum component,
but the results presented in this paper suggest a likely 
alternative interpretation.

In this paper we analyze Very Large Array (VLA) and
\emph{Chandra}  archive data of the region, mostly searching for variability
and proper  motions that could help in a better understanding of the
radio sources.

\section{Observations and Results}


The parameters of the VLA\footnote{The VLA is
a facility of the National Radio
Astronomy Observatory, which is operated by Associated Universities
Inc. under cooperative agreement with the National Science Foundation.} archive 
observations used in this study are summarized
in Table 1. All the data were calibrated following the standard VLA procedures.
The flux calibrators and the name
and bootstrapped flux
density of the phase calibrators used in each observation are given in Table 1.
In the following sections we describe in more detail the radio and X-ray observations.





\subsection{Radio Continuum Emission}


In Figure 1 we show the continuum images of the central region of W75N
at 3.6
cm (2006.38) and 2 cm (2001.31). As can be seen in this figure, to the south of the
previously known components (VLA 1, VLA 2, VLA 3 and Bc), we detected a new
radio source, VLA 4. This source was not detected in the other continuum images
discussed here (see below),
most probably because they are noiser than the 2006 image at 3.6 cm. 
In Table 2 we show the positions
of the five sources (from the 3.6 cm 2006 data) with their flux densities
at both wavelengths (3.6 and 2 cm) and their spectral indices
in this wavelength range. We note that since some of the sources
are time-variable (see below), these spectral indices are not
very reliable.

Three of the sources (VLA~1, Bc, and VLA~4) have flat spectral indices
within the error, a result consistent
with optically thin free-free emission. Source VLA~3 has a spectral index
of 0.6$\pm$0.1, consistent with the value expected for a thermal jet. The source
with the steepest spectral index is VLA~2, that with a value of 2.2$\pm$0.3 suggests
optically thick free-free emission. 


In Figure 2 we compare the three 3.6 cm images made from data taken in
1992, 1998, and 2006. Since these observations were not made with
the same phase calibrator (see Table 1), we aligned them by assuming that
the position of VLA 3 was the same at all three epochs and using the 2006 position
for all of them. 
This alignment involved small position shifts of $\leq 0\rlap.{''}1$.
In particular, the shift of the 1998 image to align it with the
2006 image was of only $\Delta \alpha = +0\rlap.^s007; \Delta \delta = -0\rlap.{''}03$.
Of the sources in the region, only source Bc shows important
changes
in position. Its average proper motion, $0\rlap.{''}3\pm 0\rlap.{''}1$ 
toward the south, over a
time of 13.48 years at a distance of 2 kpc, implies an average velocity in
the
plane of the sky of 220$\pm$70 km s$^{-1}$. However, as can be seen in Figure 2,
most of
the displacement took place between the last two epochs, so the 
velocity between these two last epochs is possibly larger.

In addition to the proper motions, there is a dramatic change in flux
density
and morphology of source Bc at 3.6 cm. 
In 1992 it was a weak
compact source (with a flux density of 0.9$\pm$0.2 mJy), becoming brighter (2.7$\pm$0.5 mJy) 
and elongated in 1998 (see Figure 2). The position does not seem to change significantly
between these two epochs.
Finally, in 2006 the source remained bright (3.3$\pm$0.2 mJy)
and appears as broken in two components. Furthermore, it has clearly moved to the
south. 


\subsection {Molecular Line Emission}

\subsubsection{SiO emission}

Using VLA archive data, we obtained the image for the J = 1--0; v = 0 
emission of the
SiO molecule (43.423858 GHz) in the W75N region shown in
the left panel of Figure 3.
The spectral line observations were made with one IF
and a total number of 63 
channels, for a velocity resolution of 1.2
km s$^{-1}$ and a velocity coverage of $\sim$80 km s$^{-1}$.
Since these observations did not
include an ad hoc bandpass calibrator, we were forced
to use as bandpass calibrators the flux (1331+305) and
phase (2012+464) calibrators.
The image is made from the velocity-integrated (moment 0)
emission. The SiO emission is compact and centered on
VLA 3, although part of the emission could be coming from VLA 2.
The SiO spectrum is shown in the right panel of Figure 3.
The observed transition is known to be a tracer of outflows
(e.g. Choi 2005).
We propose that the SiO emission could be tracing the
inner parts of an outflow associated with VLA 3,
although observations with much better angular resolution, sensitivity and
velocity coverage are needed to test this hypothesis. 

\subsubsection{NH$_3$ emission}

Also using VLA archive data, we obtained images for the (1,1) and (2,2) inversion
transitions of ammonia, at 23.694495 and 23.722633 GHz, respectively.
The spectral line observations were made with two IFs,
one centered on the (1,1) transition and the
other on the (2,2) transition. The number of channels
in each IF was 63, for a velocity resolution of
0.6 km s$^{-1}$ and a velocity coverage of $\sim$35 km s$^{-1}$.
Since these observations did not
include an ad hoc bandpass calibrator, we were forced
to use as bandpass calibrators the flux (1331+305) and
phase (2007+404) calibrators.
We made natural- and 
uniform-weight images of the NH$_3$ emission. The natural-weight images 
emphasize the large scale extended emission of the region, while the 
uniform-weight images emphasize bright and compact emission. We also made 
maps of the rotational temperature of the gas, estimated from the 
(2,2)/(1,1) ratio following Ho \& Townes (1983), assuming constant 
excitation conditions along the line of sight and optically thin emission.

In Figure 4 we show the large scale emission of the ammonia.
The main result of this image is that the region of active star formation 
in W75N appears to be located at the intersection of two molecular
filaments, one with emission in the LSR radial velocity range of
$\sim$5 to 8 km s$^{-1}$ and the other with emission in the LSR radial velocity range of
$\sim$9 to 13 km s$^{-1}$. 
In particular, the five continuum sources are located at the 
intersection zone, which is also the hottest zone. This suggests that 
the star formation in W 75N is being triggered by the collision of 
molecular filaments. This mode of star formation is predicted by 
numerical simulations (e.g., Ballesteros-Paredes et al. 1999 and 
V\'azquez-Semanedi et al. 2007) and has been also suggested
to be present for the 
regions W3 IRS 5 by Rod\'on et al. (2008) and W33A by 
Galv\'an-Madrid et al. (2010).

In Figure 5 we investigate in more detail the region of active star 
formation.
In the left panel of this Figure we show in greyscale the velocity-integrated (moment 0) 
emission of the (2,2)
transition as well as the adjacent continuum emission, obtained
from the line-free channels, in
contours.
On the right panel of this Figure we show the (1,1) and (2,2) spectra
integrated over the region of emission. From this spectra we conclude that there
is strong and broad (FWHM of $\sim$6 km s$^{-1}$)
ammonia emission originating from the region that
contains the continuum sources VLA~1, VLA~2, and VLA~3. This emission
is most probably tracing dense ($\geq 3 \times 10^4$ cm$^{-3}$) gas.
>From the ratio of the inner satellite lines to the main line of the (1,1)
transition ($\sim$0.3), we conclude that the emission is optically thin.
Following Anglada et al. (1995) and assuming a $[NH_3/H_2]$ ratio
of $10^{-7}$, we estimate a mass of $\sim$0.6 $M_\odot$ for
the condensation.

In Figure 6 we show, on the left panel, the velocity field (moment 1)
from the (2,2) transition. On the center we show
the velocity dispersion (moment 2) from the
(2,2) transition. For a Gaussian profile
the full width at half power equals the dispersion
multiplied by 2.35. Finally, on the right panel we show the rotational 
temperature of the gas.
The region with ammonia emission shows a velocity shift of
about 2 km s$^{-1}$ over a region of $\sim$3{''}.
This velocity shift could be arising from different gas  components
associated with the multiple sources (VLA1, VLA2, VLA3) within the
NH$_3$ beam. 
The central panel of Figure 6 indicates that the broadest emission overlaps
the position of the sources VLA 1, 2, and 3.
The right panel of Figure 6 indicates that the coolest parts of
the ammonia cloud, with temperatures of about 35 K,
overlaps the position of VLA~1. 

\subsection {X-ray Emission}

{\it Chandra}/ACIS level 2 event data were extracted from \emph{Chandra}
archive\footnote{See http://cda.harvard.edu/chaser/} (ObsID 8893, observed 
in 2008 February 2).
The data were reduced with the {\sc ciao 3.4}\footnote{See
http://asc.harvard.edu/ciao} data analysis system and the
\emph{Chandra}  Calibration Database (caldb 3.4.0\footnote{See 
http://cxc.harvard.edu/caldb/}).
The exposure time was processed to exclude background flares, using the 
task {\sc lc\_clean.sl}\footnote{See http://cxc.harvard.edu/ciao/download/scripts/}
in source-free sky regions of the same observation.

Although the default
pixel size of \emph{Chandra}/ACIS detector is $\sf{0.492\arcsec}$, smaller
spatial scales are accessible as the image moves across the detector
pixels during the telescope dither, thus sampling pixel scales smaller than
the default pixel of {\it Chandra}/ACIS detector. It allows us to create
0.125\arcsec~ sub-pixel binning images.
Furthermore, we applied
adaptive smoothing CIAO tool {\sc csmooth}, based on the algorithm
developed by Ebeling et al. (2006) to detect the low contrast diffuse 
emission
(minimum and maximum significance S/N level of 3 and 4,
and a scale maximum of 2 pixels).

Three energy bands images were created: 0.2-10, 0.2-2 and 2-10 keV bands.
We detected two sources;
in spite of their low count rate
($2.0 \times10^{-4}$ and $2.7 \times10^{-4}$ count/sec), it is clear
that they show emission
only above 2 keV. 
This may indicate a very obscured environment
($\rm{N_H} \geq 10^{22}$ cm$^{-2}$), although high quality data are needed in order to confirm
this possibility.  An image from the 2.0 -- 10.0 keV emission
is shown in Figure 7. The two detected sources could be  
marking the positions of termination shocks of the outflows of the
cluster of sources in W75N.
Better data are required to determine the nature of the X-ray
emission.





\section{Discussion}

\subsection{Bc: a radio HH object}

As commented in Section 2.1, the source Bc shows changes in its position, flux
density
and morphology. The source seems to be moving to the south with a velocity
larger than 200 km s$^{-1}$ while it breaks into two components and dramatically
increases its total flux density. This behavior is similar to that observed in
optical (Devine et al. 2009) and radio (Curiel et al. 1993)
Herbig-Haro objects, and, therefore, we
interpret
source Bc as an obscured radio HH object. Its radio luminosity
(3.3 mJy in 2006, at a distance of 2 kpc) makes it one of 
the brightest HH objects detected in the radio.
It is comparable to, but significantly above, HH 80 and HH 81 that have centimeter flux 
densities
of 1-2 mJy at a distance of 1.7 kpc (Mart\'\i\ et al. 1993). 
The radio luminosity of source Bc is exceeded only by the radio HH objects associated
with the luminous protostar IRAS~16547-4247, that with flux densities of
$\sim$3-4 mJy at a distance of 2.9 kpc (Rodr\'\i guez et al. 2008) are
the brightest known.

The radio flux density from source Bc allows to make an estimate of the
mechanical luminosity of the jet that drives this source. 
Assuming that the continuum emission
is optically-thin free-free at an electron temperature of $10^4$ K,
and taking into account that the source is at a distance of 2 kpc,
in the order of $\sim1.1 \times 10^{45}$ ionizations per second are needed to maintain
the source ionized. If these ionizations have a collisional origin
and assuming that a minimum energy of 13.6 eV is needed per
ionization, a mechanical luminosity of $L_{mech} \geq 6~L_\odot$ 
is estimated. Finally, assuming that the jet has a velocity of
$\sim 200$ km s$^{-1}$, a mass loss rate of order $10^{-6}~M_\odot$ yr$^{-1}$
is inferred for the jet that is producing source Bc.  

\subsection{VLA 2, VLA 3, and VLA 4}

These results lead to a revision of the nature of the radio sources in
W75N. As
commented above, Source Bc is most probably a bow shock produced by a jet.
Several arguments suggest that this jet is driven by VLA 3: i)
the source Bc and possibly VLA 4 move away from VLA 3,
and this source is elongated in the direction of motion,
ii) the spectral index is consistent with that expected for
a radio jet, and iii) VLA 3 is the only source that has
associated SiO emission (presumably tracing an outflow) and
associated 7 mm (Shepherd et al. 2004) and 1 mm 
(Shepherd 2001) continuum emission, suggesting association
with a disk or envelope.

VLA 4 is either an independent star or could be a previous ejection
from VLA 3. Unfortunately, VLA 4 is not detected
at 3.6 cm in 1992 or 1998, but a comparison of
the 3.6 cm data from 2006 and of the 2 cm data from 2001 suggests
a small displacement to the south. New sensitive observations
are needed to test this hypothesis.

Finally, VLA 2 is most probably an independent, very young star.
This source was associated with the very strong OH flare emission that
reached in 2003 an emission of about 1000 Jy to become the brightest
galactic OH maser in history (Alakoz et al. 2005; Slysh \& Migenes 2006).


\subsection{What is the Nature of VLA 1?}

Under the light of these results, one could 
propose that VLA 1 is the northern bow shock produced by the
VLA 3 jet. However, in contrast to source Bc, VLA 1 does not show
detectable proper motions along the axis of the
VLA 3 jet. This behavior could be due
to the existence of dense gas to its north that blocks 
motion along that direction.
In fact, as shown previously,
VLA 1 is associated with dense gas traced by the NH$_3$ 
observations. VLA 1 does show, however, a displacement in the peak of emission
from east to west (see Fig. 2), nearly perpendicular to the
axis of VLA 3 and similar to that observed in the radio HH objects associated
with IRAS~16547-4247
(Rodr\'\i guez et al. 2008). In this latter source,
this kinematic behavior has been interpreted as working surfaces where a precessing
jet is interacting with a very dense medium. In this case, the HH objects
would trace the present point of interaction between the jet and the dense medium.  

We think, however, most likely that VLA 1 simply traces
an independent young massive star. This source is known to be
associated with intense maser emission from several molecules: water
vapor (e.g., Lekht et al. 2009), methanol (e.g., Surcis et al. 2009),
and hydroxil (e.g. Fish \& Reid 2007). Intense maser  emission is
usually tracing the nearby presence of a young high-mass star. 
The proper motions observed in the water masers along
the major axis of the VLA 1 source also favor this scenario
(Torrelles et al. 2003).
The linear polarization of the methanol masers revealing a tightly ordered
magnetic field over more than 2000 AU around VLA 1 with a strength of
50 mG gives further support to the hypothesis that VLA 1 harbors a massive young star  
(Surcis et al. 2009).
It is relevant to note that the flux density of VLA 1 remained constant
within the noise during the three epochs of 3.6 cm observations, 
with all values consistent with 4.0$\pm$0.2 mJy.

\section{Conclusions}

We have analyzed VLA and \emph{Chandra} archive data of 
the W75N region. In the radio continuum, we detect
five sources: VLA1, VLA2, 
VLA3, Bc, and VLA4, from north to south respectively. Our main conclusions follow:

1) We detect important changes in total flux density, morphology, and position
in the source Bc, suggesting that it is not tracing an
independent star but actually is a radio HH object powered by VLA~3. Its average velocity
in the plane of the sky is 220$\pm$70 km s$^{-1}$. If our interpretation is
correct, this is one of the brightest radio HH objects known.

2) We detect J = 1--0; v = 0 emission of SiO centered on
VLA 3, probably tracing the
inner parts of an outflow associated with this radio jet.
Our results strongly support the identification of VLA 3 as
a radio jet.

3) The large scale ammonia emission shows that the W75N region contains 
two filamentary molecular clouds with different radial velocities. The star 
formation is taking place at the intersection of these two filaments, 
which is also the hottest region. This suggests that the star formation 
could be triggered by the collision of the two filamentary clouds.

4) Strong and broad (1,1) and (2,2) ammonia emission
is detected from the region containing the radio sources
VLA~1, VLA~2, and VLA~3. 

5) We detected a new radio continuum component, VLA 4, a few arcsec to the south of
the group of previously known sources.

6) Two sources are detected in the 2-10 keV band \emph{Chandra}/ACIS image. These two 
sources could be tracing the termination shocks 
of outflows in the region, but better data
are needed to understand the nature of the X-ray emission. 




\acknowledgments

L.F.R. is thankful for the support
of DGAPA, UNAM, and of CONACyT (M\'exico).
G.A., C.C.-G., and J.M.T. acknowledge support from MICINN (Spain) 
AYA2008-06189-C03 grant (co-funded with FEDER funds), and from Junta de 
Andaluc\'\i a (Spain).



{\it Facilities:} \facility{VLA, \emph{Chandra}}.








\clearpage



\begin{figure}
\epsscale{1.0}
\plotone{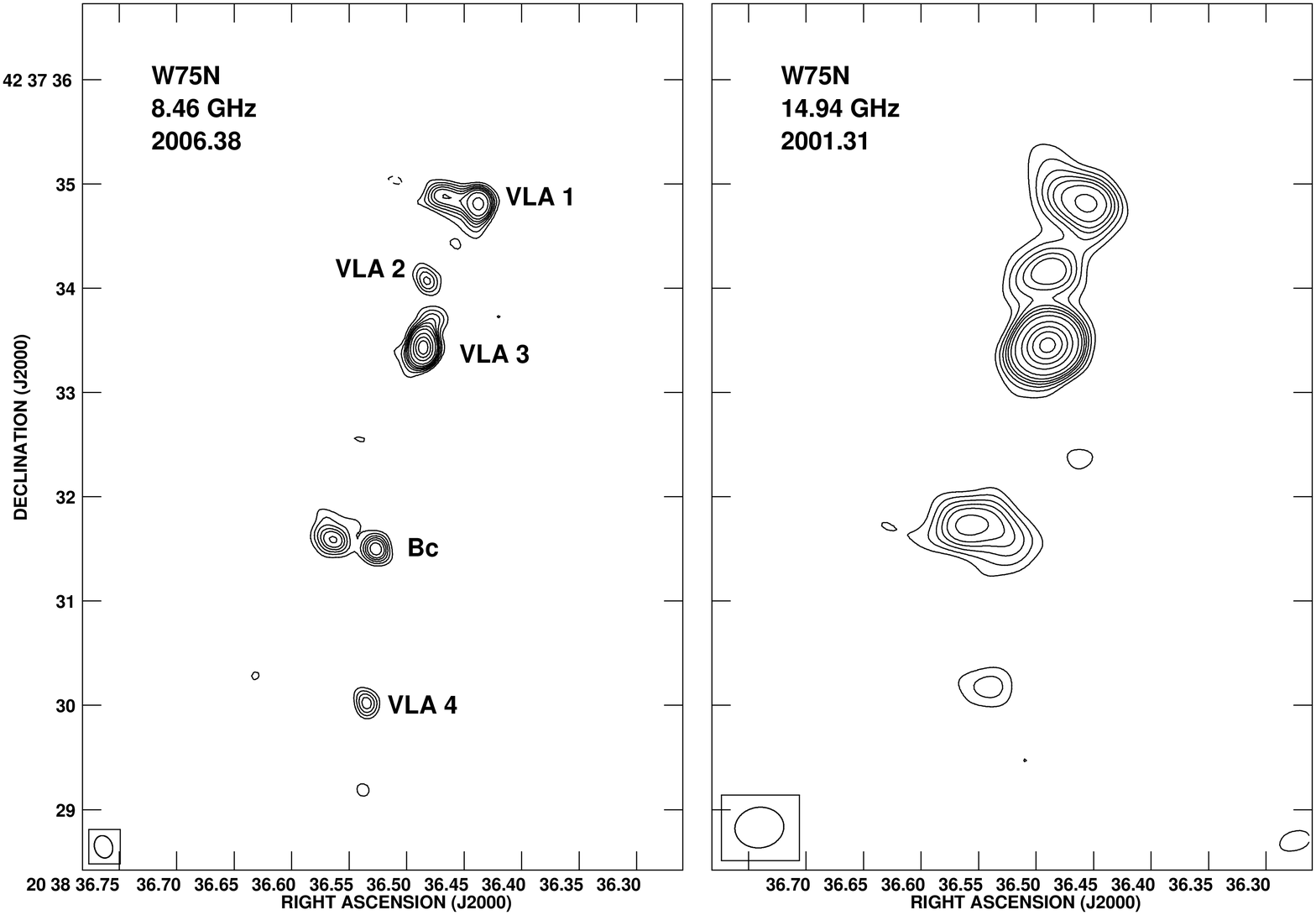}
\caption{VLA contour images of the 3.6 cm (left) and 2 cm (right) emission
from the W75N region. The contours are -4, 4, 6, 8, 10, 12, 15, 20, 30, 40, 50, and 60 times
45 and 78 $\mu$Jy beam$^{-1}$, the respective rms noises. 
The half power contours of the synthesized beams are shown in the bottom left
corner of each figure and are $0\rlap.{''}22 \times 0\rlap.{''}17;
PA = 18^\circ$ for the 3.6 cm image
and $0\rlap.{''}47 \times 0\rlap.{''}39; PA = -83^\circ$ for the 2 cm image. 
Both images were made with
ROBUST = 0 (u,v) weighting. The source VLA 4 is a new detection. 
The epochs of the observations are given in the figure.
\label{fig1}
}
\end{figure}

\clearpage




\begin{figure}
\plotone{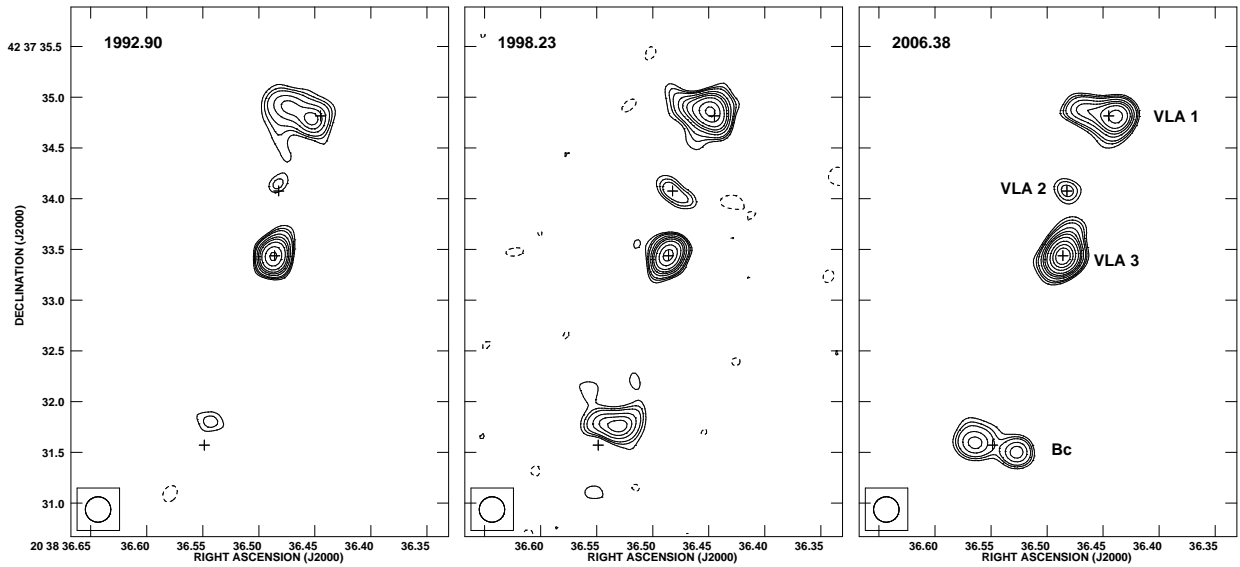}
\caption{VLA contour images of the 3.6 cm emission for the three epochs
(1992.90, 1998.23, and 2006.38) analyzed.
Contours are  -4, 4, 5, 6, 8, 10, 12, 15, 20, 25, and 30
times 80 $\mu$Jy beam$^{-1}$. The images have been
restored with a circular beam of $0\rlap.{''}25$,  
shown in the bottom left corner of the images. The crosses mark the centroids of
the radio sources for epoch 2006.38.
The source VLA 4 is detected only in the last epoch and it is not included
in these images.
\label{fig2}}
\end{figure}

\begin{figure}
\plotone{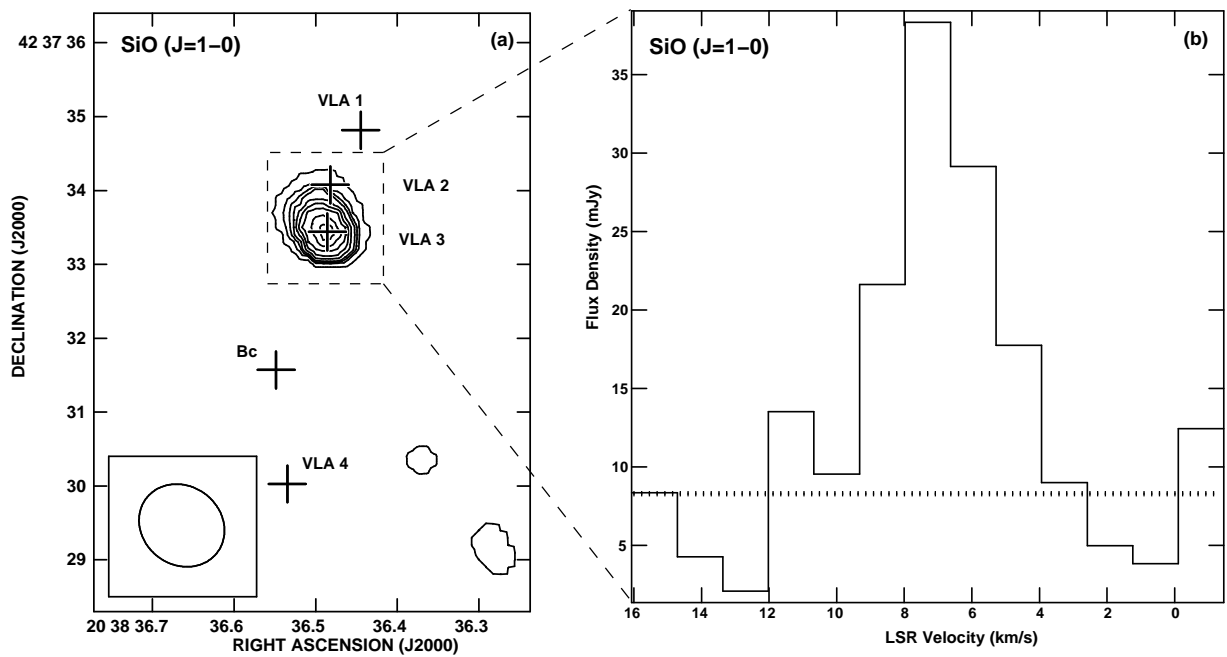}
\caption{a) VLA contour image of the
moment 0 of the J= 1--0; v = 0 SiO emission from the W75N region.
Contours are
20, 40, 50, 60, 80, 100, 120, 140, and 160 mJy beam$^{-1}$ km s$^{-1}$.
The beam ($1\rlap.{''}21 \times 1\rlap.{''}06$; PA = $52^\circ$)
is shown in the bottom left corner of the image. The crosses mark the positions
of the radio continuum sources. The dashed rectangle indicates the solid angle
over which the spectrum of the right panel of this
figure was obtained. b) Spectrum of the SiO emission from the box indicated
in the left panel. The dotted horizontal line indicates the expected
7-mm continuum contribution, as estimated from the results of Shepherd et al. (2004). 
\label{fig3}}
\end{figure}

\begin{figure}
\plotone{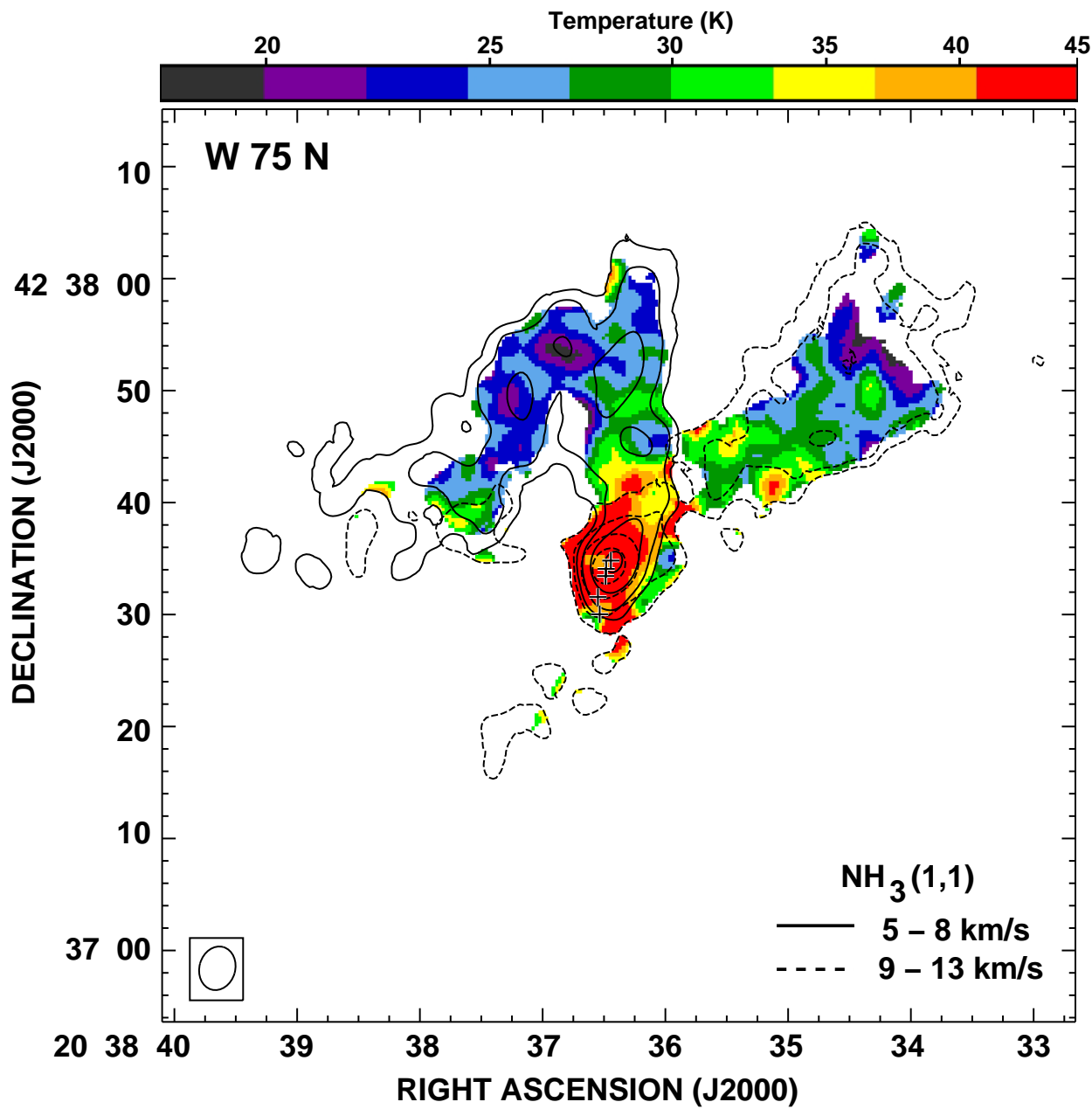}
\caption{Superposition of the natural-weight images of the moment 0 of the NH$_3$(1,1)
emission (contours) over 
the temperature image (colors). The solid contours trace the emission integrated in the 
velocity range from 5 to 8 km~s$^{-1}$, while the dashed contours are 
integrated in the velocity range from 9 to 13 km~s$^{-1}$. Contour 
levels are 30, 50, 100 and 200 mJy~beam$^{-1}$~km~s$^{-1}$. The color scale
indicates the rotational temperature, as derived
from the (2,2)/(1,1) ratio. The crosses 
mark the positions of the five continuum sources.
The beam ($4\rlap.{''}1 \times 3\rlap.{''}$; PA = $-18^\circ$)
is shown in the bottom left corner of the image.
\label{fig4}}
\end{figure}

\begin{figure}
\plotone{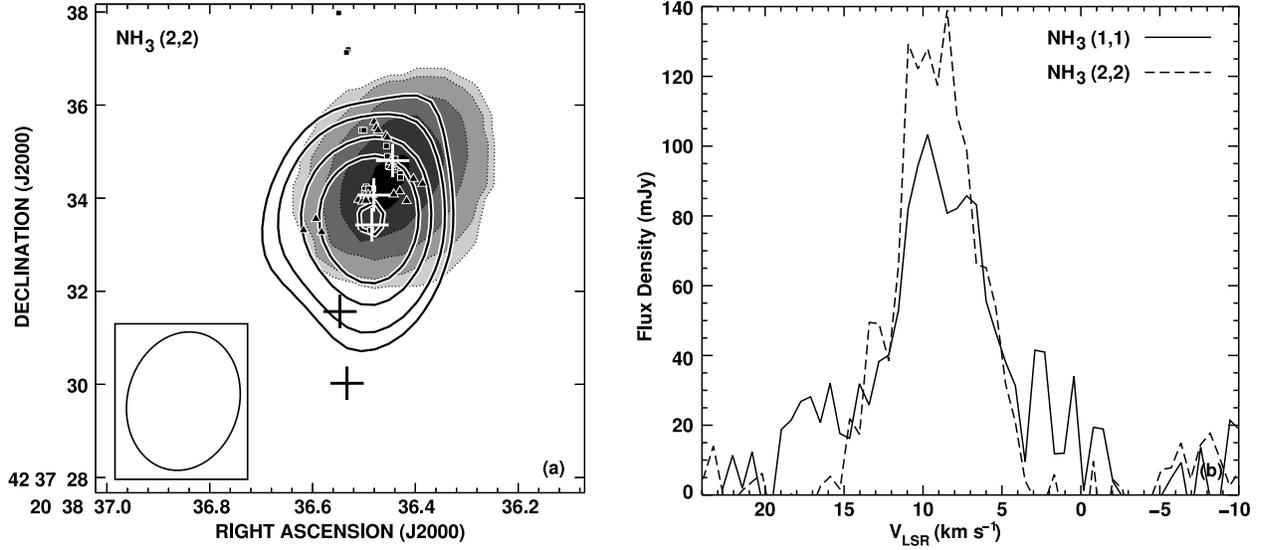}
\caption{a) In greyscale levels we show the velocity-integrated (moment 0)
emission from the (2,2)
transition of ammonia. Levels are 50, 100, 200, 300, and 400 mJy beam$^{-1}$ km s$^{-1}$.
The thin contours show the
1.3 cm emission continuum from the region
(obtained from the line-free channels). Contours are 
-3, 3, 4, 6, 8, and 12 times 1.7 mJy beam$^{-1}$, the rms of the
continuum image.
The five crosses mark the positions of the continuum sources.
The squares are the positions of the H$_2$O masers from
Torrelles et al. (1997) and the triangles the 
OH masers from Hutawarakorn et al. (2002).
The beam ($2\rlap.{''}97 \times 2\rlap.{''}43$; PA = $-13^\circ$)
is shown in the bottom left corner.
b) Spectra of the (1,1) and (2,2) transitions of ammonia
integrated over the region of emission.
Lines are broad, and in the (1,1) 
spectrum the emission of the main line appears blended with that of the 
inner satellites.
\label{fig5}}
\end{figure}

\begin{figure}
\plotone{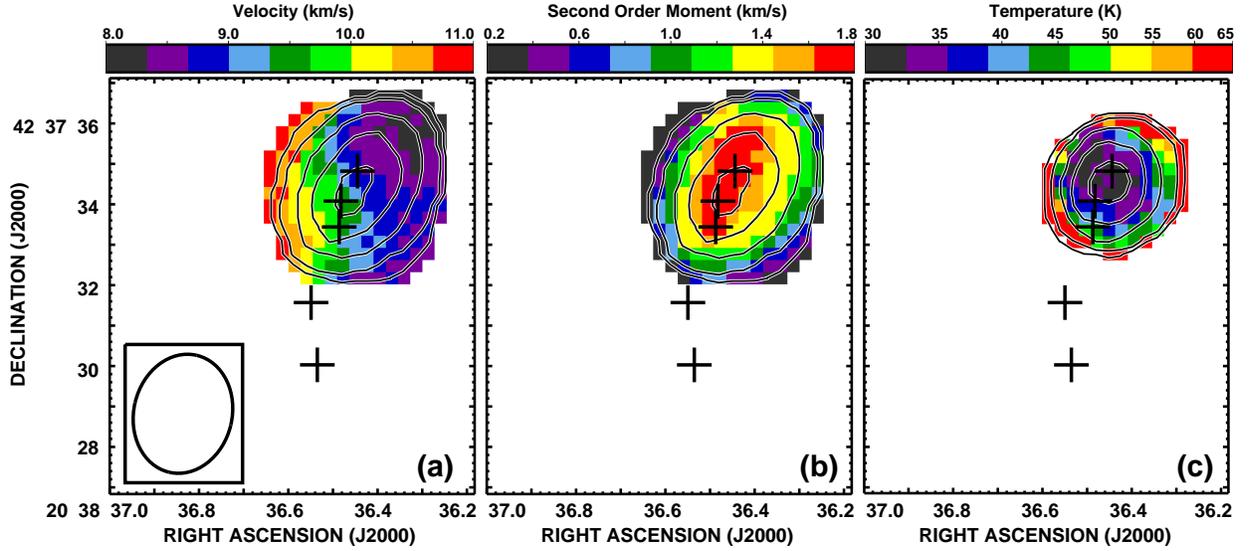}
\caption{a) Velocity field (moment 1)
from the (2,2) transition of ammonia. The
color wedge at the top gives the LSR radial velocity in
km s$^{-1}$. The beam ($2\rlap.{''}97 \times 2\rlap.{''}43$; PA = $-13^\circ$)
is shown in the bottom left corner.
In all panels the contours mark the integrated emission (moment 0)
of the (2,2) emission, contours are 50, 100, 200, 300, 
and 400 mJy beam$^{-1}$ km s$^{-1}$.
The crosses mark the positions of the five radio sources.
b) Velocity dispersion (moment 2)
of the (2,2) emission.
For a Gaussian line the FWHM is 2.35 times the value of the velocity 
dispersion.
c) Rotational
temperature of the gas, estimated from the (2,2)/(1,1) ratio,
assuming optically thin emission 
and constant excitation conditions along the line of sight.
The color wedge at the top gives the temperature in K.
\label{fig6}}
\end{figure}

\begin{figure}
\plotone{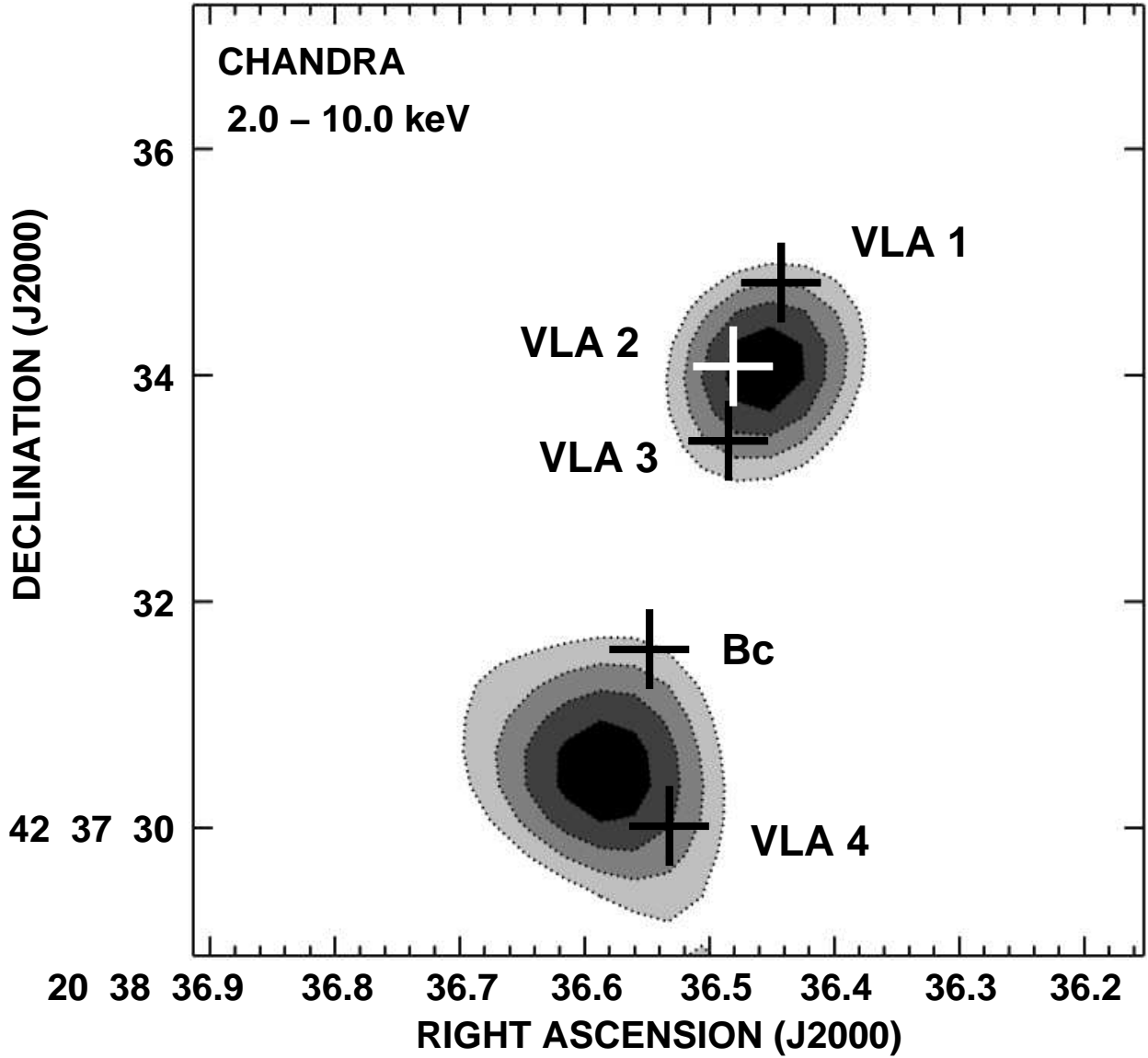}
\caption{Left: \emph{Chandra}/ACIS greyscale image of the 2.0 - 10.0 keV
emission from the W75N region. Succesive greyscale levels
are 0.05, 0.06, 0.07, and 0.08 counts/pixel. The crosses mark the positions
of the five radio sources. 
\label{fig7}}
\end{figure}







\clearpage
\small

\begin{deluxetable}{cccccccccc}
\tabletypesize{\scriptsize}
\rotate
\tablewidth{0pt}
\tablenum{1}
\tablecolumns{10}
\tablecaption{Parameters of the VLA observations.\label{tbl-1}}
\tablehead{ \colhead{} & \colhead{}  & \colhead{Configuration/} & \colhead{Frequency} & 
\colhead{Flux} & \colhead{Phase Calibrator/} & \colhead{On-source} & 
  \colhead{Calibration} & \colhead{Syntesized} & \colhead{RMS} \\ 
 \colhead{Project} & \colhead{Epoch} & \colhead{Mode} & \colhead{(GHz)} & 
 \colhead{Calibrator} & \colhead{Bootstrapped Flux Density (Jy)} & \colhead{Time (min)} & 
    \colhead{Cycle (min)} & \colhead{Beam$^a$} & \colhead{(mJy)}} 
\startdata
 AT141 & 1992 Nov 24 (1992.90) & A/Continuum & 8.44 & 0137+331 & 2007+404/3.18$\pm$0.07 & 5.3 &
 8 & $0\rlap.{''}23 \times 0\rlap.{''}19; -56^\circ$ & 0.09 \\
 AA224 & 1998 Mar 26 (1998.23) & A/Continuum & 8.46 & 1331+305 & 2025+337/4.73$\pm$0.08 & 2.3 &
 4 & $0\rlap.{''}24 \times 0\rlap.{''}20; 48^\circ$ & 0.13 \\
 AS678 & 2000 Apr 24 (2000.31) & C/Line & 43.4 & 1331+305 & 2012+464/1.07$\pm$0.02 & 196.4 &
 2 & $1\rlap.{''}21 \times 1\rlap.{''}06 ;  52^\circ$ & 3.2 \\
 AM652 & 2000 Aug 03 (2000.59) & D/Line & 23.7 & 1331+305 & 2007+404/2.30$\pm$0.03 & 51.5 &
 15 & $2\rlap.{''}97 \times 2\rlap.{''}43 ; -13^\circ$ & 6.6 \\
 AF381 & 2001 Apr 23 (2001.31) & B/Continuum & 14.9 & 1331+305 & 2015+371/2.21$\pm$0.01 & 68.5 &
 4  & $0\rlap.{''}47 \times 0\rlap.{''}39 ; -83^\circ$ & 0.08 \\
 AS831 & 2006 May 18 (2006.38) & A/Continuum & 8.46 & 1331+305 & 2007+404/2.30$\pm$0.01 & 50.7 &
 12 & $0\rlap.{''}22 \times 0\rlap.{''}17 ;  18^\circ$ & 0.05 \\
\enddata
\tablenotetext{a}{Major axis$\times$minor axis; position angle.}
\end{deluxetable}



\clearpage

\begin{table}
\begin{center}
\caption{Parameters of the continuum sources.\label{tbl-2}}
\small
\vskip0.2cm
\begin{tabular}{ccccccc}
\tableline\tableline
 &  &  & Deconvolved Size$^a$  & $S_\nu$(3.6 cm)$^b$ & $S_\nu$(2 cm)$^b$ & Spectral \\
Source & $\alpha$(2000)$^a$ & $\delta$(2000)$^a$ & ($'' \times '';~^\circ$) & (mJy) & 
(mJy) & Index$^c$  \\
\tableline
VLA 1 & $20^h~ 38^m~ 36\rlap.^s445$ & $+42^\circ~ 37'~ 34\rlap.{''}82$ & 
0.43$\pm$0.02$\times$0.17$\pm$0.02; 80$\pm$2 &
3.8$\pm$0.1 & 3.0$\pm$0.2 & -0.4$\pm$0.1 \\
VLA 2 & $20^h~ 38^m~ 36\rlap.^s482$ & $+42^\circ~ 37'~ 34\rlap.{''}08$ &
$< 0\rlap.{''}2$ &
0.7$\pm$0.1 & 2.4$\pm$0.3 & 2.2$\pm$0.3 \\
VLA 3 & $20^h~ 38^m~ 36\rlap.^s486$ & $+42^\circ~ 37'~ 33\rlap.{''}44$ &
0.21$\pm$0.01$\times$0.07$\pm$0.01; 157$\pm$3 &
4.0$\pm$0.1 & 5.7$\pm$0.2 & 0.6$\pm$0.1 \\
Bc & $20^h~ 38^m~ 36\rlap.^s549$ & $+42^\circ~ 37'~ 31\rlap.{''}57$ &
0.75$\pm$0.05$\times$0.23$\pm$0.03; 74$\pm$3 &
3.3$\pm$0.2 & 3.0$\pm$0.3 & -0.2$\pm$0.2 \\
VLA 4 & $20^h~ 38^m~ 36\rlap.^s535$ & $+42^\circ~ 37'~ 30\rlap.{''}03$ &
$< 0\rlap.{''}2$ &
0.7$\pm$0.1 & 0.9$\pm$0.2 & 0.4$\pm$0.5 \\
\tableline
\end{tabular}
\tablenotetext{a}{From the 3.6 cm 2006 data. Absolute positional error is estimated to be
$\sim 0\rlap.{''}07$. The deconvolved dimensions are given as
major axis $\times$ minor axis; position angle.}
\tablenotetext{b}{The flux densities were obtained fitting a single Gaussian ellipsoid
to all the emission present. This fitting was made with the task JMFIT
of AIPS and the error was estimated using the rms of the image and
the theory described in Beltr\'an et al. (2001).}
\tablenotetext{c}{Spectral index in the 
3.6 to 2 cm wavelength range.}
\end{center}
\end{table}





\begin{thebibliography}{}

\bibitem[Alakoz et al.(2005)]{2005AstL...31..375A} Alakoz, A.~V., Slysh, 
V.~I., Popov, M.~V., \& Val'Tts, I.~E.\ 2005, Astronomy Letters, 31, 375 

\bibitem[Anglada et al.(1995)]{1995ApJ...443..682A} Anglada, G., Estalella, 
R., Mauersberger, R., Torrelles, J.~M., Rodriguez, L.~F., Canto, J., Ho, 
P.~T.~P., \& D'Alessio, P.\ 1995, \apj, 443, 682 

\bibitem[Baart et al.(1986)]{1986MNRAS.219..145B} Baart, E.~E., Cohen, 
R.~J., Davies, R.~D., Rowland, P.~R., 
\& Norris, R.~P.\ 1986, \mnras, 219, 145 

\bibitem[BHV99]{BHV99} Ballesteros-Paredes, J., Hartmann, L. \& V\'azquez-Semanedi, E. 
1999, ApJ, 527, 285

\bibitem[Beltr{\'a}n et al.(2001)]{2001AJ....121.1556B} Beltr{\'a}n, M.~T., 
Estalella, R., Anglada, G., Rodr{\'{\i}}guez, L.~F., 
\& Torrelles, J.~M.\ 2001, \aj, 121, 1556 

\bibitem[Carrasco-Gonz{\'a}lez et 
al.(2006)]{2006A&A...445L..43C} Carrasco-Gonz{\'a}lez, C., L{\'o}pez, R., 
Gyulbudaghian, A., Anglada, G., \& Lee, C.~W.\ 2006, \aap, 445, L43 

\bibitem[Choi(2005)]{2005ApJ...630..976C} Choi, M.\ 2005, \apj, 630, 976 

\bibitem[Curiel et al.(1993)]{1993ApJ...415..191C} Curiel, S., Rodriguez, 
L.~F., Moran, J.~M., \& Canto, J.\ 1993, \apj, 415, 191 


\bibitem[Devine et al.(2009)]{2009AJ....137.3993D} Devine, D., Bally, J., 
Chiriboga, D., \& Smart, K.\ 2009, \aj, 137, 3993 

\bibitem[Eb(06]{Ebeling} Ebeling, H., White, D.A., \& Rangarajan, F.V.N. 2006, MNRAS, 368, 65

\bibitem[Eisloffel et al.(2000)]{2000prpl.conf..815E} Eisloffel, J., Mundt, 
R., Ray, T.~P., \& Rodriguez, L.~F.\ 2000, Protostars and Planets IV, 815 

\bibitem[Feigelson 
\& Montmerle(1999)]{1999ARA&A..37..363F} Feigelson, E.~D., \& Montmerle, T.\ 1999, \araa, 37, 363 

\bibitem[Fischer et al.(1985)]{1985ApJ...293..508F} Fischer, J., Sanders, 
D.~B., Simon, M., \& Solomon, P.~M.\ 1985, \apj, 293, 508 

\bibitem[Fish 
\& Reid(2007)]{2007ApJ...656..952F} Fish, V.~L., \& Reid, M.~J.\ 2007, \apj, 656, 952 

\bibitem[Galv{\'a}n-Madrid et al.(2010)]{2010,in.preparation} 
Galv{\'a}n-Madrid, R., et al.\ 2010, in preparation

\bibitem[Garay et al.(1996)]{1996ApJ...459..193G} Garay, G., Ram\'\i rez, S., 
Rodr\'\i guez, L.~F., Curiel, S., \& Torrelles, J.~M.\ 1996, \apj, 459, 193 

\bibitem[Haschick et al.(1981)]{1981ApJ...244...76H} Haschick, A.~D., Reid, 
M.~J., Burke, B.~F., Moran, J.~M., \& Miller, G.\ 1981, \apj, 244, 76 

\bibitem[Henriksen et 
al.(1991)]{1991A&A...248..221H} Henriksen, R.~N., Mirabel, I.~F., \& 
Ptuskin, V.~S.\ 1991, \aap, 248, 221 

\bibitem[Ho 
\& Townes(1983)]{1983ARA&A..21..239H} Ho, P.~T.~P., \& Townes, C.~H.\ 1983, \araa, 21, 239 

\bibitem[Hunter et 
al.(1994)]{1994A&A...284..215H} Hunter, T.~R., Taylor, G.~B., Felli, M., \& Tofani, G.\ 
1994, \aap, 284, 215

\bibitem[Hutawarakorn et al.(2002)]{2002MNRAS.330..349H} Hutawarakorn, B., 
Cohen, R.~J., \& Brebner, G.~C.\ 2002, \mnras, 330, 349 

\bibitem[Lekht et al.(2009)]{2009ARep...53..420L} Lekht, E.~E., Slysh, 
V.~I., \& Krasnov, V.~V.\ 2009, Astronomy Reports, 53, 420 

\bibitem[Marti et al.(1993)]{1993ApJ...416..208M} Marti, J., Rodriguez, 
L.~F., \& Reipurth, B.\ 1993, \apj, 416, 208 

\bibitem[O'Dell 
\& Wong(1996)]{1996AJ....111..846O} O'Dell, C.~R., \& Wong, K.\ 1996, \aj, 111, 846 

\bibitem[Persi et 
al.(2006)]{2006A&A...445..971P} Persi, P., Tapia, M., \& Smith, H.~A.\ 2006, \aap, 445, 971 

\bibitem[RBM08]{RBM08} Rod\'on, J.A., Beuther, H., Megeath, S.T., \& van der Tak, 
F.F.S. 2008, A\&A, 490, 213

\bibitem[Rodr{\'{\i}}guez et al.(2008)]{2008AJ....135.2370R} 
Rodr{\'{\i}}guez, L.~F., Moran, J.~M., Franco-Hern{\'a}ndez, R., Garay, G., 
Brooks, K.~J., \& Mardones, D.\ 2008, \aj, 135, 2370 

\bibitem[Shepherd(2001)]{2001ApJ...546..345S} Shepherd, D.~S.\ 2001, \apj, 
546, 345 

\bibitem[Shepherd et al.(2003)]{2003ApJ...584..882S} Shepherd, D.~S., 
Testi, L., \& Stark, D.~P.\ 2003, \apj, 584, 882 

\bibitem[Shepherd et al.(2004)]{2004ApJ...601..952S} Shepherd, D.~S., 
Kurtz, S.~E., \& Testi, L.\ 2004, \apj, 601, 952 

\bibitem[Slysh 
\& Migenes(2006)]{2006MNRAS.369.1497S} Slysh, V.~I., \& Migenes, V.\ 2006, \mnras, 369, 1497 

\bibitem[Surcis et 
al.(2009)]{2009A&A...506..757S} Surcis, G., Vlemmings, W.~H.~T., Dodson, R., \& van 
Langevelde, H.~J.\ 2009, \aap, 506, 757 

\bibitem[Torrelles et al.(1997)]{1997ApJ...489..744T} Torrelles, J.~M., 
G\'omez, J.~F., Rodr\'\i guez, L.~F., Ho, P.~T.~P., Curiel, S., 
\& V\'azquez, R.\ 1997, \apj, 489, 744 

\bibitem[Torrelles et al.(2003)]{2003ApJ...598L.115T} Torrelles, J.~M., et 
al.\ 2003, \apjl, 598, L115 

\bibitem[VG07]{VG07} V\'azquez-Semanedi, E., G\'omez, G.C., Jappsen, A.K., 
Ballesteros-Paredes, J., Gonz\'alez, R.F., \& Klessen, R.S. 2007, ApJ, 
657, 870

\bibitem[Zapata et al.(2004)]{2004AJ....127.2252Z} Zapata, L.~A., 
Rodr{\'{\i}}guez, L.~F., Kurtz, S.~E., 
\& O'Dell, C.~R.\ 2004, \aj, 127, 2252 


\end{thebibliography}
\end{document}